\documentclass[12pt,preprint]{aastex}

\newcommand{\mlangle}{\langle}
\newcommand{\mrangle}{\rangle}

\DeclareSymbolFont{bsymbols}{OML}{cmm}{bx}{it}
\DeclareSymbolFontAlphabet{\mathbf}{bsymbols}

\renewcommand{\vec}{\mathbf}

\def\imag{{\rm i}}

\def\ket#1{|{#1}\mrangle\kern .175em}
\def\bra#1{\kern .175em\mlangle{#1}|}

\def\<{\kern-5pt}

\def\threej#1#2#3#4#5#6{\left(\matrix{#1&\<#2&\<#3 \cr
				 #4&\<#5&\<#6 \cr}\right)}
\def\sixj#1#2#3#4#5#6{\left\{\matrix{#1&\<#2&\<#3 \cr
				 #4&\<#5&\<#6 \cr}\right\}}
\def\ninej#1#2#3#4#5#6#7#8#9{\left\{\matrix{#1&\<#2&\<#3 \cr
				 #4&\<#5&\<#6 \cr
				 #7&\<#8&\<#9 \cr}\right\}}

\begin{document}

\title{On the atomic polarization of the ground level of 
\ion{N\lowercase{a}}{1}}

\author{Roberto Casini}
\affil{High Altitude Observatory,
National Center for Atmospheric Research,\altaffilmark{1}
P.O.~Box 3000, Boulder CO 80307-3000}
\author{Egidio Landi Degl'Innocenti}
\affil{Dipartimento di Astronomia e Scienza dello Spazio,
Universit\`a di Firenze, Largo E.~Fermi 2, I-50125 Firenze, Italy}
\author{Marco Landolfi}
\affil{Istituto Nazionale di Astrofisica, Osservatorio Astrofisico 
di Arcetri, Largo E.~Fermi 5, I-50125 Firenze, Italy}
\and
\author{Javier Trujillo Bueno\altaffilmark{2}}
\affil{Instituto de Astrof{\'\i}sica de Canarias, V{\'\i}a L\'actea s/n,
E-38200 La Laguna, Tenerife, Spain}

\altaffiltext{1}{The National Center for Atmospheric Research is
sponsored by the National Science Foundation}

\altaffiltext{2}{Consejo Superior de Investigaciones Cient\'\i ficas, 
Spain}

\begin{abstract}

In a recent letter (\citealt{TB02}), we showed the remarkable 
result that the atomic alignment of the levels P$_{1/2}$ and 
S$_{1/2}$ of the D$_1$ line of \ion{Na}{1} 
is practically destroyed in the presence of magnetic fields 
sensibly larger than $10\,\rm G$, irrespectively of the field 
direction. In this paper, we demonstrate analytically 
that this property is a consequence
of the decoupling of the electronic and nuclear angular momenta,
$\vec J$ and $\vec I$, in the excited state P$_{3/2}$, which is
achieved when the Zeeman splitting from the local magnetic field
becomes much larger than the typical hyperfine separation for 
that level.

\end{abstract}

\subjectheadings{atomic processes --- polarization --- 
scattering --- Sun: magnetic fields}

\section{Introduction}
\label{sec:intro}

The observation and theoretical modeling of weak polarization 
signatures in spectral lines are opening a new window on the
investigation of the weak magnetism of the 
solar atmosphere (see, e.g., 
the recent reviews by \citealt{STEN01,TB01a,TBMA02}). 
To this aim, it is important to investigate carefully 
within the framework of the quantum theory of polarization
(e.g., Landi Degl'Innocenti 1983) the observable effects of the 
atomic polarization of the energy levels involved in the line 
transitions of interest, including their subtle modification 
by the presence of magnetic fields. 

In this respect, in a recent letter (\citealt{TB02}, hereafter 
Paper~I), we reported on an interesting property of the 
polarizability of the levels of the D$_1$ line of \ion{Na}{1}: 
in spite of the fact that those levels can both be 
aligned,\footnote{\label{note:atompol} Atomic 
{\em alignment} is a condition of population imbalances between 
the Zeeman substates of a level, such that the total populations
of substates with different values of $|M|$ are different. One 
speaks instead of atomic {\em orientation} when, for a given 
value of $|M|$, the substates labeled by $M$ and $-M$ have 
different populations. See, e.g., \citet{LA84}, or the
recent review by \citet{TB01a}.} when proper account is 
taken of the additional quantum numbers introduced by the 
hyperfine structure (HFS) of \ion{Na}{1}, the alignment is
drastically reduced for fields larger than $10\,\rm G$, and
practically vanishes for $B\ga 100\,\rm G$, {\it irrespective} 
of the relative directions of the magnetic field and of the
incident radiation. Accordingly, any contribution to the
linear polarization in the core of D$_1$ that arises from 
atomic alignment is suppressed for magnetic fields sensibly larger
than $10\,\rm G$, so the only expected linear-polarization signal 
for such field strengths must be due to the transverse Zeeman effect 
(see Fig.~2 of Paper~I; the reader should note how the Stokes-$Q$ 
signature of single-scattering events taking place in the presence 
of a vertical magnetic field changes from antisymmetric
for $B<10\,\rm G$ to symmetric for $B\ga 50\,\rm G$).

In Paper~I, we were concerned mainly with a detailed calculation
of the polarizability of the \ion{Na}{1} levels, and with the 
consequences it bears for our understanding of the magnetic-field 
distribution 
and topology in the solar atmosphere. In the present work, we 
focus instead on the investigation of the atomic physics that 
is behind the polarization properties of those lines.

To this end, we follow our approach of Paper~I, and apply the 
quantum theory of line formation 
in the limit of complete frequency redistribution (CRD)
and in the collisionless regime, as 
developed by \citet{LA83,LA84,LA85}, to investigate the 
statistical equilibrium (SE) 
of an ensemble of \ion{Na}{1} atoms illuminated by anisotropic
radiation (see also \citealt{LL85}). The hypothesis of CRD 
corresponds to the requirement that the incident radiation field 
coming from the underlying photosphere, and illuminating the 
scattering atoms, be spectrally flat over an interval much larger 
than the energy separation between atomic levels whose wavefunctions 
sensibly overlap (leading to the phenomenon of quantum interferences).
In the case of the D$_1$ and D$_2$ lines forming in the 
solar atmosphere, this is a good assumption only if we neglect the 
quantum interferences between the upper levels of D$_1$ and 
D$_2$. More specifically, these are interferences between the 
levels P$_{1/2}$ and P$_{3/2}$ pertaining to the same atomic 
term. Whereas the role of these so-called {\it super-interferences}
is important for a correct interpretation of line polarization
in the wings of D$_1$ and D$_2$, the line-core polarization of 
those lines, which was the subject of the investigation of Paper~I,
is expected to be largely unaffected by them.

In \S~2, we summarize our qualitative description
of the polarization properties of the levels of \ion{Na}{1}
(see Paper~I),
and introduce some useful new concepts and terminology.
In \S~\ref{sec:polar.quant}, we put those concepts 
on a more quantitative basis, and provide an algebraic proof
that the alignment of the levels of D$_1$ is suppressed when a
magnetic regime of complete decoupling of the angular momenta 
$\vec J$ and $\vec I$ is reached in the excited state P$_{3/2}$.
Finally, in the conclusive section, we provide further arguments
to illuminate this interesting phenomenon.

\section{Polarizability of the \ion{Na}{1} levels: qualitative
description}
\label{sec:polar.qual}

The stable isotope of sodium has a nuclear spin $I=3/2$, therefore
we must take into account the role of HFS in the solution of the 
SE problem of \ion{Na}{1}. HFS was already indicated by \citet{LA98}
as the only possible mechanism  allowing for the existence of atomic 
alignment in the levels of the D$_1$ line.
In fact, levels with 
total angular momentum $J=1/2$ cannot be aligned, whereas both 
hyperfine levels $F=1$ and $F=2$, into which a level $J=1/2$ splits 
in the coupling process with a nuclear spin $I=3/2$, can be aligned.

For this reason, it is convenient to introduce the concept of 
{\it intrinsic polarizability} (IP), for those levels whose values 
of $J$ allow the presence of atomic alignment, and of 
{\it extrinsic polarizability} (EP), for those levels
that can carry atomic alignment only through the ``internal''
$F$ quantum numbers, because of the presence of HFS. (What 
distinguishes the roles of $J$ and $F$ as quantum numbers, in 
this context, is the assumption we made at the beginning, that 
quantum interferences can exist only between different $F$ 
levels, but not between different $J$ levels.) In this sense, 
we can speak of EP only in the cases of $J=0$ and $J=1/2$.
Therefore, 
both levels of the D$_1$ line of
\ion{Na}{1} are EP, whereas the upper level of D$_2$ is IP, 
because $J=3/2$. 

\begin{figure*}[t]
\epsscale{0.8}
\plotone{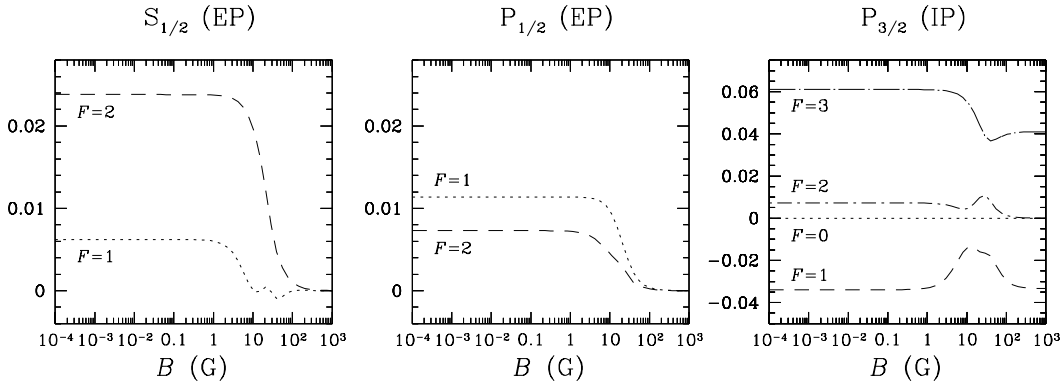}
\figcaption{%
\label{fig:alignment}
Fractional atomic alignment,
$\sigma^2_0(F)=\rho^2_0(F,F)/\rho^0_0(F,F)$, of the three lowest levels 
of \ion{Na}{1} against the magnetic field strength. A vertical field, 
and a height of 10" of the scattering atoms above the solar surface,
are assumed. The kinetic temperature of the emitting plasma is 
$T=6000\,\rm K$. These results show that the atomic alignment of 
the levels of D$_1$ is practically zero 
when $B\ga 100\,\rm G$, even for vertical fields.}
\end{figure*}

This nomenclature has a direct link with the physics of the 
interaction processes of the atom with the incident radiation field. 
We speak of IP of an atomic level when this level has the 
possibility of absorbing the multipole order $K=2$ of 
the polarization tensor of the incident radiation field 
(\citealt{LA83}; see also \citealt{TB01a}), 
expressed in the irreducible spherical tensor 
representation, $J^K_Q$ ($Q=-K,\ldots,K$). In particular,
if we assume that the incident radiation field is unpolarized,
and has cylindrical symmetry around the local solar vertical 
through the scattering centre, only the multipole orders $K=0$ 
(intensity) and $K=2$ (anisotropy) are present in the 
radiation-field tensor. In this case, it is found that an EP 
level can only absorb the multipole order $K=0$, so there is 
no atomic polarization directly induced by the
incident radiation field. Any atomic 
alignment ($K=2$, in the irreducible spherical representation of 
the density matrix) that such level can show---when proper
account is taken of its sub-structure associated with HFS---can 
only come from the transfer of atomic alignment from
other atomic levels that are instead IP. In the case of \ion{Na}{1}, 
for instance, if the two levels of D$_1$ were isolated (i.e., 
not radiatively connected with other levels in the atom), 
no atomic alignment could be created, even accounting for 
the presence of HFS. Because of the presence of the upper level 
P$_{3/2}$ of D$_2$ in the SE problem of \ion{Na}{1}, instead, transfer 
of atomic polarization from such IP level to the lower level of D$_1$ 
can occur, via the radiative de-excitation associated with the 
formation of the D$_2$ line. Once EP has been created in the level 
S$_{1/2}$, this can be transferred via absorption processes to the 
upper level of D$_1$ as well.
In our case, the two levels of D$_1$ manifest their EP because
of the alignment induced onto the corresponding HFS levels, with 
$F=1,2$ (see Fig.~1 of Paper~I; also Fig.~\ref{fig:alignment}
introduced below).

On the other hand, the transfer of atomic alignment from 
an IP level to an EP level can be inhibited under particular
conditions. For the three-level model of the \ion{Na}{1} atom 
considered here, and for the prescribed radiation field,
we determined that  the atomic polarization 
in the two EP levels vanishes when the IP level P$_{3/2}$ reaches 
the regime of the complete Paschen-Back effect, in which the 
Zeeman splittings of the $F$ levels due to the local magnetic 
field become much larger than the HFS separations between those 
levels. In this regime, the HFS coupling of the electronic and 
nuclear angular momenta, $\vec J$ and $\vec I$, of the \ion{Na}{1} 
atoms in the excited state P$_{3/2}$, is ``relaxed'' by the presence 
of the strong magnetic field, through the electronic Zeeman
effect.\footnote{The nuclear Zeeman effect is completely negligible
in our picture, up to field strengths of the order of $10^5\,\rm G$.}
[To understand the meaning of such decoupling process, we must
observe that, in the regime of complete Paschen-Back effect, and 
assuming the direction of $\vec B$ as the quantization axis, 
$J_z$ becomes a conserved quantity (rigorously, an element of the
complete set of commuting observables of the atomic system), along
with $F_z$. Because $I_z=F_z-J_z$ must be conserved as well, both 
$M_J$ and $M_I$ become good quantum numbers, so the eigenvectors of 
the atomic system take the form $\ket{JM_J,IM_I}$.]


\begin{figure*}[t]
\epsscale{0.8}
\plotone{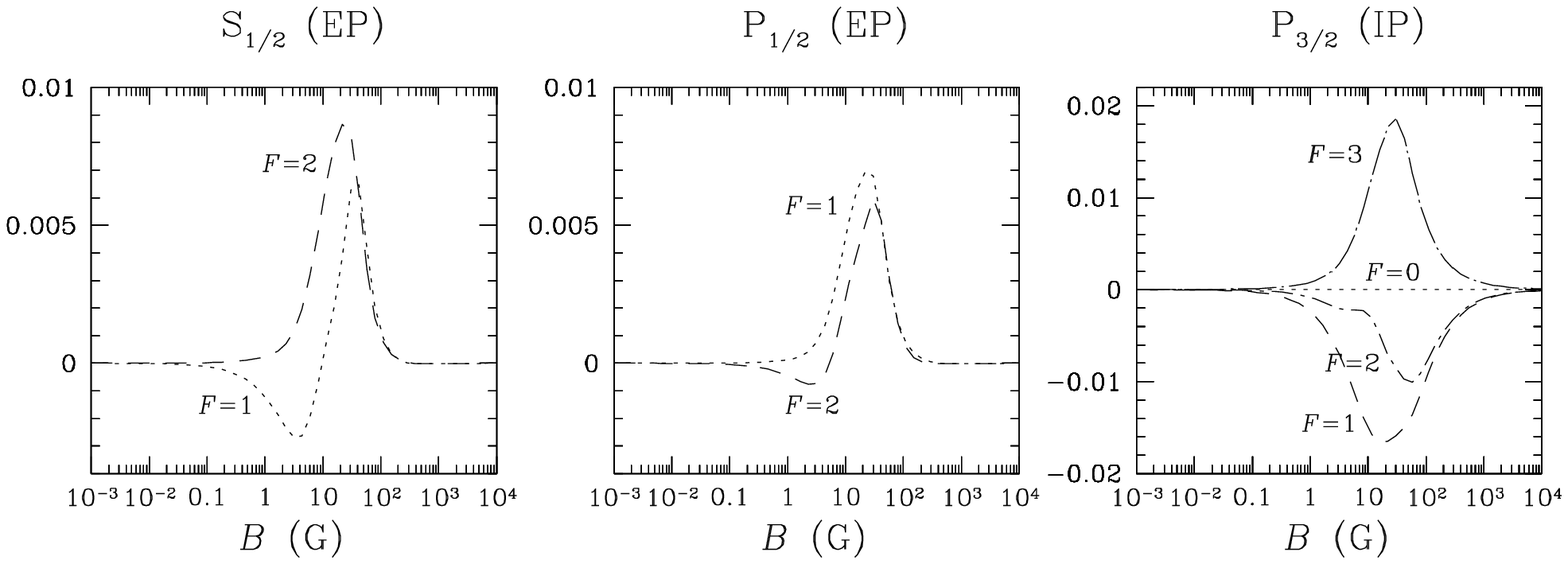}
\figcaption{%
\label{fig:orientation}
Fractional atomic orientation,
$\sigma^1_0(F)=\rho^1_0(F,F)/\rho^0_0(F,F)$, of the three lowest 
levels of \ion{Na}{1} against the magnetic field strength (notice 
the different range of magnetic strengths with respect to
Fig.~\ref{fig:alignment}). The 
scattering geometry and the plasma conditions are the same as in
the calculation of Figure~\ref{fig:alignment}. These results show 
that atomic orientation in the levels of D$_1$
is negligible for fields $B\ga 100\,\rm G$.}
\end{figure*}


The inhibition of the transfer of atomic alignment from an IP
level to an EP level for increasing magnetic strengths is clearly 
illustrated by the results 
presented in Paper~I. In Figure~\ref{fig:alignment}, we reproduce similar
results. We calculated the atomic alignment of the levels of 
D$_1$ and D$_2$ for magnetic strengths between $10^{-4}\,\rm G$
and $10^3\,\rm G$.
A vertical field (i.e., aligned along the symmetry axis of the 
radiation cone from the photosphere illuminating the scattering 
atom) was chosen, in order to clarify that the obtained trend of the 
alignment against the magnetic field strength is {\it not} due 
to Hanle-effect depolarization. As we see, atomic alignment in 
the levels of D$_1$ is drastically reduced for fields 
larger than $10\,\rm G$, and practically vanishes for fields of 
the order of $100\,\rm G$ or larger. In Figure~\ref{fig:orientation}, 
we show analogous results for the atomic orientation ($K=1$, in the 
irreducible spherical representation of the density matrix; see
Note~\ref{note:atompol} for a description of atomic orientation), 
for magnetic strengths between $10^{-3}\,\rm G$ and $10^4\,\rm G$. 
Also in this case, the orientation of the levels of the D$_1$ line 
practically vanishes for $B\ga 100\,\rm G$. (We note that,
for $B>100\,\rm G$, the level P$_{3/2}$ approaches the regime of 
complete Paschen-Back effect. In fact, for $B\sim 500\,\rm G$, the 
typical Zeeman splitting is already 10 times larger than the typical
HFS separation for that level.)

On the other hand, as suggested by the work of \citet{LE69} concerning
optical-pumping processes in cadmium, a sufficient condition for 
the vanishing of atomic alignment in the EP level is that the HFS 
frequency separation of the IP level be negligible with respect to the 
Einstein $A$-coefficient of the atomic transition. This condition 
is very general, as it holds regardless of the magnetic-field 
strength (in particular, it is valid also for zero magnetic fields). 
In the case of \ion{Na}{1}, the HFS frequency separation of the 
IP level P$_{3/2}$ is comparable with the Einstein $A$-coefficient of 
the D$_2$ line.\footnote{Coincidentally, this implies that the regime of 
complete Paschen-Back effect in the upper level also corresponds to 
the saturation regime of the Hanle effect for that level.}
For this reason, transfer of atomic alignment from the 
IP upper level to the EP lower level is possible when $B < 100\,\rm
G$, since $\vec J$ and $\vec I$ are still significantly coupled in 
the IP level P$_{3/2}$ (cf.~also Paper~I, end of \S 3).

These results suggest that the inhibition of the transfer of atomic 
alignment from an IP level to an EP level should be regarded
as an aspect of the so-called {\it principle of spectroscopic
stability} applied to the IP level: Whenever the hyperfine
structure of the IP level becomes negligible, whether because a 
magnetic field is present which is strong enough to reach the 
complete Paschen-Back regime for that level, or because the HFS 
separation of that level is much smaller than its radiative width, 
the transfer of alignment from the IP level to the EP level is 
inhibited, so the EP level behaves as if the atomic HFS were not 
present at all. The reason for this is hidden in the complexity of 
the SE problem, and it is addressed in the following section.

\section{Polarizability of the \ion{Na}{1} levels: analytical
description}
\label{sec:polar.quant}

We consider an IP level with total electronic angular momentum $J$. 
We assume that this level can only interact with EP levels in the atom.
Beyond this restriction, the atomic system can be arbitrary, so the
following formalism applies also for atoms other than \ion{Na}{1}.
If a nuclear spin is present, of angular momentum $I$, the density 
matrix for the IP level in the irreducible spherical tensor 
representation is (cf.~\citealt{LA84})
\begin{equation}
\label{eq:rhoKQ}
{}^{JI}\rho^K_Q(F,F')
	=\sum_{M_F M_F'}(-1)^{F-M_F}\sqrt{\mathstrut 2K+1}\,
	\threej{F}{F'}{K}{M_F}{-M_F'}{-Q}\,
	{}^{JI}\rho(FM_F,F'M_F')\;.
\end{equation}
We write explicitly
\begin{equation}
\label{eq:rhoFF}
{}^{JI}\rho(FM_F,F'M_F')=\bra{(JI)FM_F}\rho\ket{(JI)F'M_F'}\;,
\end{equation}
where
\begin{equation}
\label{eq:ket}
\ket{(JI)FM_F}=\sum_{M_J M_I}C(JM_J\,IM_I;FM_F)\ket{JM_J,IM_I}\;.
\end{equation}
In the previous equation, $C(JM_J\,IM_I;FM_F)$ are Clebsh-Gordan
coefficients, which can be expressed in terms of $3j$ symbols as
\begin{equation}
\label{eq:CG}
C(JM_J\,IM_I;FM_F)=(-1)^{J-I+M_F}\sqrt{\mathstrut 2F+1}\,
	\threej{J}{I}{F}{M_J}{M_I}{-M_F}.
\end{equation}
Substitution of eq.~(\ref{eq:ket}) into eq.~(\ref{eq:rhoFF}),
using eq.~(\ref{eq:CG}), gives
\begin{eqnarray}
\label{eq:rhoFF.1}
{}^{JI}\rho(FM_F,F'M_F') \vphantom{\sum}
&=&(-1)^{M_F-M_F'}\,\sqrt{(2F+1)(2F'+1)} \nonumber \\
&&\kern -30mm 
	\times\sum_{M_J M_I M_J'M_I'}
	\threej{J}{I}{F}{M_J}{M_I}{-M_F}
	\threej{J}{I}{F'}{M_J'}{M_I'}{-M_F'}
	\bra{JM_J,IM_I}\rho\ket{JM_J',IM_I'}\;.
\end{eqnarray}

We now make the assumption that the electronic spin and the 
nuclear spin are decoupled (or very weakly coupled) when the 
atom is in the IP level. As anticipated in the previous section, 
this can be the case if the HFS separation is much smaller than 
the natural width of that level, or, in the presence of a magnetic 
field, if the level is in the regime of complete Paschen-Back effect. 
In either case, the atomic density matrix for the IP level can 
be factorized as
\begin{equation}
\label{eq:rho.fact}
\bra{JM_J,IM_I}\rho\ket{JM_J',IM_I'}
	=\rho(JM_J,JM_J')\,\rho(IM_I,IM_I')\;.
\end{equation}
We introduce at this point the formalism of the irreducible spherical 
tensors for both $\rho(JM_J,JM_J')$ and $\rho(IM_I,IM_I')$,
\begin{mathletters}
\begin{eqnarray}
\label{eq:rhoJ}
\rho(JM_J,JM_J')
&=&\sum_{K_J Q_J}(-1)^{J-M_J}\sqrt{\mathstrut 2K_J+1}\,
	\threej{J}{J}{K_J}{M_J}{-M_J'}{-Q_J} \rho^{K_J}_{Q_J}(J)\;, \\
\label{eq:rhoI}
\rho(IM_I,IM_I')
&=&\sum_{K_I Q_I}(-1)^{I-M_I}\sqrt{\mathstrut 2K_I+1}\,
	\threej{I}{I}{K_I}{M_I}{-M_I'}{-Q_I} \rho^{K_I}_{Q_I}(I)\;.
\end{eqnarray}
\end{mathletters}
Substitution of eq.~(\ref{eq:rho.fact}) into eq.~(\ref{eq:rhoFF.1}), 
using eqs.~(\ref{eq:rhoJ}) and (\ref{eq:rhoI}), gives
\begin{eqnarray*}
{}^{JI}\rho(FM_F,F'M_F')
&=&\vphantom{\sum}
	(-1)^{J+I-M_F'}\,\sqrt{(2F+1)(2F'+1)}
	\sum_{K_J Q_J}\sum_{K_I Q_I} \sqrt{(2K_J+1)(2K_I+1)}\,
	\rho^{K_J}_{Q_J}(J)\,\rho^{K_I}_{Q_I}(I) \nonumber \\
&&\kern -38mm 
	\times \sum_{M_J M_J'M_I M_I'}
	\threej{J}{I}{F}{M_J}{M_I}{-M_F}
	\threej{J}{I}{F'}{M_J'}{M_I'}{-M_F'}
	\threej{J}{J}{K_J}{M_J}{-M_J'}{-Q_J}
	\threej{I}{I}{K_I}{M_I}{-M_I'}{-Q_I}.
\end{eqnarray*}
Finally, this equation must be substituted into eq.~(\ref{eq:rhoKQ}). 
We then obtain an expression which involves the contraction over
all magnetic quantum numbers of a product of five $3j$ symbols.
This contraction can be evaluated using, e.g., eq.~(14), p.~456,
of \citet{VA88}, yielding the expression
\begin{eqnarray}
\label{eq:rhoKQ.JI}
{}^{JI}\rho^K_Q(F,F')
&=&(-1)^{K-Q}\,\sqrt{(2K+1)(2F+1)(2F'+1)}
	\sum_{K_J K_I} \sqrt{(2K_J+1)(2K_I+1)}\,
	\ninej{J}{I}{F}{J}{I}{F'}{K_J}{K_I}{K} \nonumber \\
&&\times \sum_{Q_J Q_I} \threej{K}{K_J}{K_I}{Q}{-Q_J}{-Q_I}
	\rho^{K_J}_{Q_J}(J)\,\rho^{K_I}_{Q_I}(I)\;.
\end{eqnarray}

As a particular case, if nuclear polarization is absent 
($K_I=Q_I=0$), eq.~(\ref{eq:rhoKQ.JI}) reduces to
\begin{equation}
\label{eq:rhoKQ.JI.0}
{}^{JI}\rho^K_Q(F,F')=(-1)^{J+I+F'+K}\,\rho^0_0(I)\,
	\sqrt{(2F+1)(2F'+1)\over 2I+1}\,
	\sixj{J}{J}{K}{F}{F'}{I} \rho^K_Q(J)\;.
\end{equation}
In this case, the (electronic) atomic polarization of the $J$ 
level translates {\it directly} (i.e., with the same $K$ and $Q$) 
into the atomic polarization of the $(F,F')$ pair. 

\subsection{The effect of very weak $J$-$I$ coupling on the 
SE problem}

As an application of the former development, we consider a 
two-level atom $(J_u,J_l)$ endowed with HFS. Neglecting 
stimulated emission for simplicity, the SE equations for the 
two levels read (\citealt{LA83,LA84,LA85})
\begin{eqnarray}
\label{eq:SEu}
{d\over dt}\,{}^{J_u I}\rho^{K_u}_{Q_u}(F_u,F_u')
&=&-\imag \sum_{F_u''F_u'''}\sum_{K_u'Q_u'}
	{}^{J_u I}\rho^{K_u'}_{Q_u'}(F_u'',F_u''')\,
	N(F_u F_u'K_u Q_u;F_u''F_u'''K_u'Q_u') \nonumber \\
&&-\sum_{F_u''F_u'''}\sum_{K_u'Q_u'}
	{}^{J_u I}\rho^{K_u'}_{Q_u'}(F_u'',F_u''')\,
	R_{\rm E}(F_u F_u'K_u Q_u;F_u''F_u'''K_u'Q_u') \nonumber \\
&&+\sum_{F_l F_l'}\sum_{K_l Q_l}
	{}^{J_l I}\rho^{K_l}_{Q_l}(F_l,F_l')\,
	T_{\rm A}(F_u F_u'K_u Q_u;F_l F_l'K_l Q_l)
\end{eqnarray}
and
\begin{eqnarray}
\label{eq:SEl}
{d\over dt}\,{}^{J_l I}\rho^{K_l}_{Q_l}(F_l,F_l')
&=&-\imag \sum_{F_l''F_l'''}\sum_{K_l'Q_l'}
	{}^{J_l I}\rho^{K_l'}_{Q_l'}(F_l'',F_l''')\,
	N(F_l F_l'K_l Q_l;F_l''F_l'''K_l'Q_l') \nonumber \\
&&-\sum_{F_l''F_l'''}\sum_{K_l'Q_l'}
	{}^{J_l I}\rho^{K_l'}_{Q_l'}(F_l'',F_l''')\,
	R_{\rm A}(F_l F_l'K_l Q_l;F_l''F_l'''K_l'Q_l') \nonumber \\
&&+\sum_{F_u F_u'}\sum_{K_u Q_u}
	{}^{J_u I}\rho^{K_u}_{Q_u}(F_u,F_u')\,
	T_{\rm E}(F_l F_l'K_l Q_l;F_u F_u'K_u Q_u)\;.
\end{eqnarray}

To understand how atomic polarization is created in an EP level,
assuming that the other level is IP, we must consider explicitly
the expressions of the transfer rates for absorption and 
spontaneous emission processes, respectively,
\begin{eqnarray}
\label{eq:TA}
T_{\rm A}(F_u F_u'K_u Q_u;F_l F_l'K_l Q_l)
&=&(2J_l+1)\,B_{J_l J_u}
	\sqrt{(2F_u+1)(2F_u'+1)} \nonumber \\
&&
	\times \sum_{K_r Q_r}\sqrt{3(2K_u+1)(2K_l+1)(2K_r+1)}\,
	\threej{K_u}{K_l}{K_r}{-Q_u}{Q_l}{-Q_r}
	J^{K_r}_{Q_r}(\omega_{ul}) \nonumber \\
&&\kern -5cm 
	\times (-1)^{F_l'-F_l+K_l+Q_l}\,
	\sqrt{\mathstrut\smash{(2F_l+1)(2F_l'+1)}}\,
	\ninej{F_u}{F_l}{1}{F_u'}{F_l'}{1}{K_u}{K_l}{K_r}
	\sixj{J_u}{J_l}{1}{F_l}{F_u}{I}
	\sixj{J_u}{J_l}{1}{F_l'}{F_u'}{I}
\end{eqnarray}
and
\begin{eqnarray}
\label{eq:TE}
T_{\rm E}(F_l F_l'K_l Q_l;F_u F_u'K_u Q_u)
&=&\delta_{K_l K_u}\,\delta_{Q_l Q_u}\,
	(2J_u+1)\,A_{J_u J_l}\,
	\sqrt{\mathstrut\smash{(2F_l+1)(2F_l'+1)}} \nonumber \\
&&\kern -5cm 
	\times(-1)^{F_l'+F_u'+K_l+1}\sqrt{(2F_u+1)(2F_u'+1)}\,
	\sixj{F_l}{F_l'}{K_l}{F_u'}{F_u}{1}
	\sixj{J_u}{J_l}{1}{F_l}{F_u}{I}
	\sixj{J_u}{J_l}{1}{F_l'}{F_u'}{I}.
\end{eqnarray}
The relaxation rate due to spontaneous emission, $R_{\rm E}$, 
is completely diagonal, so it can only relate each of the
elements ${}^{JI}\rho^K_Q(F,F')$ to itself. The relaxation 
rate due to absorption, $R_{\rm A}$, is a necessary ingredient
of this demonstration. However, the only fact we will rely
upon is the presence in that rate of the $6j$ symbol
\begin{equation}
\label{eq:6j.RA}
\sixj{J_l}{J_l}{K_r}{1}{1}{J_u}.
\end{equation}

The rate $N$, in both eqs.~(\ref{eq:SEu}) and
(\ref{eq:SEl}), describes magnetic and HFS depolarization.
The importance of this rate is that it accounts for the 
conversion mechanism of atomic alignment ($K=2$) into 
atomic orientation ($K=1$) discussed by \citet{KE84}. 
This is related to the fact that, 
in the algebraic expression of the rate (not given here), 
$K_u$ and $K_u'$ (cf.~eq.~[\ref{eq:SEu}]) or $K_l$ and $K_l'$ 
(cf.~eq.~[\ref{eq:SEl}]) can have different parity. If the 
radiation illuminating the atom is not circularly polarized
(which is the case of the present discussion), 
this conversion mechanism is the only process capable of 
creating orientation in the atomic system (see, e.g.,
\citealt{LL85}). On the other
hand, this mechanism is only effective when quantum 
interferences between different $F$ levels are important, 
which corresponds to a regime of magnetic fields such that
level crossing between $F$ levels can occur. Therefore, for 
magnetic fields such that the upper level approaches
the regime of complete Paschen-Back effect ($B>100\,\rm G$), 
the conversion of atomic alignment into atomic orientation is 
drastically reduced (see Fig.~\ref{fig:orientation}). For this 
reason, the role of the rate $N$ is not of immediate concern for 
the following arguments.

We first consider the case in which $J_u$ is the IP level.
When this level is in a regime of very weak coupling between
$\vec J$ and $\vec I$ (whether because the HFS separation is
much smaller than $A_{J_u J_l}$, or because a magnetic field
is present that is strong enough to establish a regime of
complete Paschen-Back effect in that level), the 
irreducible components of the density matrix for that level, 
${}^{J_u I}\rho^{K_u}_{Q_u}(F_u,F_u')$, can be written according 
to eq.~(\ref{eq:rhoKQ.JI}). It is then found that the double summation 
over $F_u$ and $F_u'$ in eq.~(\ref{eq:SEl}) can be performed 
algebraically. This corresponds to a contracted product of a 
$9j$ symbol with three $6j$ symbols, which is evaluated using, 
e.g., eq.~(36), p.~471, of \citet{VA88}. The result is that 
the overall contribution of the transfer rate $T_{\rm E}$ to 
eq.~(\ref{eq:SEl}) is proportional to the product (notice that
$K_l=K_u$)
\begin{equation}
\label{eq:contr1}
\ninej{J_l}{I}{F_l}{J_l}{I}{F_l'}{K_{J_u}}{K_I}{K_l}
\sixj{J_l}{J_l}{K_{J_u}}{J_u}{J_u}{1}.
\end{equation}
Since $J_l<1$ for the EP level, the former product vanishes
unless $K_{J_u}<2$. In particular, to create alignment
in the EP lower level ($K_l=2$), either both electronic and
nuclear orientations ($K_{J_u}=1,K_I=1,3$) or only nuclear 
alignment ($K_{J_u}=0,K_I=2$) must be present when the
atom is in the excited state $J_u$. 

To convince ourselves that these conditions cannot be met, 
let us assume that initially (i.e., before irradiation) 
atomic polarization is completely absent, in particular 
$K_l=0$. Since the level $J_l$ is EP, it is only sensitive 
(through the relaxation rate $R_{\rm A}$; cf.~the $6j$ symbol
[\ref{eq:6j.RA}]) to the intensity of the incident radiation 
field, so lower-level polarization ($K_l>0$) cannot be 
created directly by irradiation. Therefore, when irradiation 
begins, from eqs.~(\ref{eq:SEu}) and (\ref{eq:TA}) we see that the 
prescribed radiation field ($K_r=0,2$) can only induce atomic 
alignment in the upper level (besides populating it),
because of the selection rule introduced by the $3j$ symbol
in eq.~(\ref{eq:TA}). Since the atom was initially unpolarized, 
and by assumption the electronic and nuclear systems are decoupled 
in the excited state, $J_u$, the atomic alignment of the upper level
can only be electronic. In fact, electric-dipole transitions cannot 
affect the nuclear system, so the nuclear Zeeman sublevels remain 
naturally populated in all 
cases of interest, even if strong $J$-$I$ coupling is present in the 
EP level. From this argument, we conclude that $K_{J_u}=0,2$, and
$K_I=0$, as a result of the excitation process. As anticipated above,
we can dismiss the alignment-to-orientation conversion mechanism as
a possible source of upper-level orientation ($K_{J_u}=1$), because 
of the assumed regime of weak $J$-$I$ coupling. Also, upper-level
alignment ($K_{J_u}=2$) cannot be transferred in the de-excitation
process, because the product (\ref{eq:contr1}) vanishes. 
Therefore, nuclear polarization can never be created in this regime, 
and eq.~(\ref{eq:rhoKQ.JI.0}) applies to the upper level. Under these 
conditions, the product (\ref{eq:contr1}) vanishes identically for 
$K_l>0$, so lower-level polarization cannot be created. This is in
agreement with the results of Paper~I, and of Figures~1 and 2 in this
paper.

In summary, when the IP upper level is in a regime of very weak 
$J$-$I$ coupling, the creation of atomic alignment in the EP lower 
level ($K_l=2$) by transfer of atomic alignment from the IP upper 
level ($K_{J_u}=2$) is inhibited. In the case of \ion{Na}{1},
this implies that the ground level $S_{1/2}$ cannot be aligned,
and consequently also the upper level P$_{1/2}$ of D$_1$ must
have zero alignment, as illustrated in Paper~I and by 
Figure~\ref{fig:alignment} in this paper. Lower-level orientation 
($K_l=1$) can in principle be created directly by irradiation, 
if $J_l=1/2$, although it requires that the incident radiation be 
circularly polarized ($K_r=1$; cf.~the $6j$ symbol [\ref{eq:6j.RA}]). 
In our case, because of the prescribed radiation field, lower-level
orientation can only be created by the transfer of atomic 
orientation from the upper level ($K_{J_u}=1$), which is not
inhibited in principle. On the other 
hand, the alignment-to-orientation conversion mechanism in the 
upper level becomes very inefficient for very weak $J$-$I$ coupling 
(see Fig.~\ref{fig:orientation}), so also upper-level orientation 
can only be created if the incident radiation field is circularly 
polarized.

We checked our conclusion that eq.~(\ref{eq:rhoKQ.JI.0}) must
apply to the IP upper level, in the regime of very weak $J$-$I$ 
coupling, against the numerical results of Paper~I (cf.~also
Fig.~\ref{fig:alignment} in this paper). In particular, we verified
that the ratio of the quantities $\sigma^2_0(3)$ and $\sigma^2_0(1)$
for the upper level $\hbox{P}_{3/2}$ of \ion{Na}{1} ($I=3/2$) in
the strong-field limit ($B=1000\,\rm G$; see rightmost panels of
Fig.~1 in Paper~I; also Fig.~\ref{fig:alignment} in this paper)
is correctly reproduced by eq.~(9). This equation also accounts
for the curious vanishing of the quantity $\sigma^2_0(2)$ in the same
limit, which is due to the fact that $\rho^2_Q(2,2)$ vanishes
identically because of the (non-trivial) nullity of the $6j$ symbol 
in eq.~(\ref{eq:rhoKQ.JI.0}). 

Within the same approximation of the two-level atom $(J_u,J_l)$, 
we now assume that the decoupling of $\vec J$ and $\vec I$ is reached 
first in the lower level, while strong coupling is still present 
in the upper level. This time we assume that $J_l$ is the IP level, 
whereas $J_u$ is the EP level. 
Since we assumed that the lower level is in a regime of very weak
coupling between $\vec J$ and $\vec I$, the irreducible 
components of the density matrix for that level, 
${}^{J_l I}\rho^{K_l}_{Q_l}(F_l,F_l')$, can be written according 
to eq.~(\ref{eq:rhoKQ.JI}). It is then found that the double 
summation over $F_l$ and $F_l'$ in eq.~(\ref{eq:SEu}) can be 
performed algebraically. This corresponds to a contracted product 
of two $9j$ symbols with two $6j$ symbols that is evaluated using, 
e.g., eq.~(37), p.~471, of \citet{VA88}. The result is that the 
overall contribution of the transfer rate $T_{\rm A}$ to 
eq.~(\ref{eq:SEu}) is proportional to the sum
\begin{equation}
\label{eq:contr2}
\sum_k (-1)^k\,(2k+1)
	\ninej{1}{J_l}{J_u}{1}{J_l}{J_u}{K_r}{K_{J_l}}{k}
	\ninej{I}{F_u}{J_u}{I}{F_u'}{J_u}{K_I}{K_u}{k}
	\sixj{K_r}{K_{J_l}}{k}{K_I}{K_u}{K_l}.
\end{equation}
Since $J_u<1$ for the EP level, this sum is limited to $k=0,1$.
Again, we assume that the atomic polarization is absent before
irradiation. Because the lower level is IP, lower-level alignment 
can be created when irradiation begins. However, since $\vec J$ and
$\vec I$ are decoupled in the lower level, nuclear polarization
remains zero ($K_I=0$), so all the alignment of the lower level 
must be electronic ($K_{J_l}=2$). Under these conditions, the sum
(\ref{eq:contr2}) is restricted to $k=0$ only, because the first $9j$
symbol in the sum (\ref{eq:contr2}) vanishes for $k=1$ unless 
$K_r+K_{J_l}$ is an odd integer. Therefore atomic polarization in 
the upper level 
($K_u>0$) can never be created, because of nullity of the second 
$9j$ symbol in the sum (\ref{eq:contr2}).

This shows, in particular, that the concept of EP is also valid for 
an upper level. In this case, the EP upper level is sensitive to 
the anisotropy of radiation ($K_r=2$) through the transfer rate 
$T_{\rm A}$, but nonetheless creation of alignment in the upper 
level through the absorption of that anisotropy is not possible 
when the IP lower level is in a regime of very weak $J$-$I$ coupling,
because of the selection rules implied by the sum (\ref{eq:contr2}). 
Upper-level orientation ($K_u=1$) is not excluded in principle, 
if $J_u=1/2$, although it can only be created by transfer of atomic 
orientation from the lower level ($K_{J_l}=1$; see
eq.~[\ref{eq:contr2}]). However, when the lower
level is in the regime of weak $J$-$I$ coupling, its orientation
can only be due to the presence of circular polarization in the 
incident radiation field. 

%


\begin{figure*}[t]
\epsscale{0.8}
\plotone{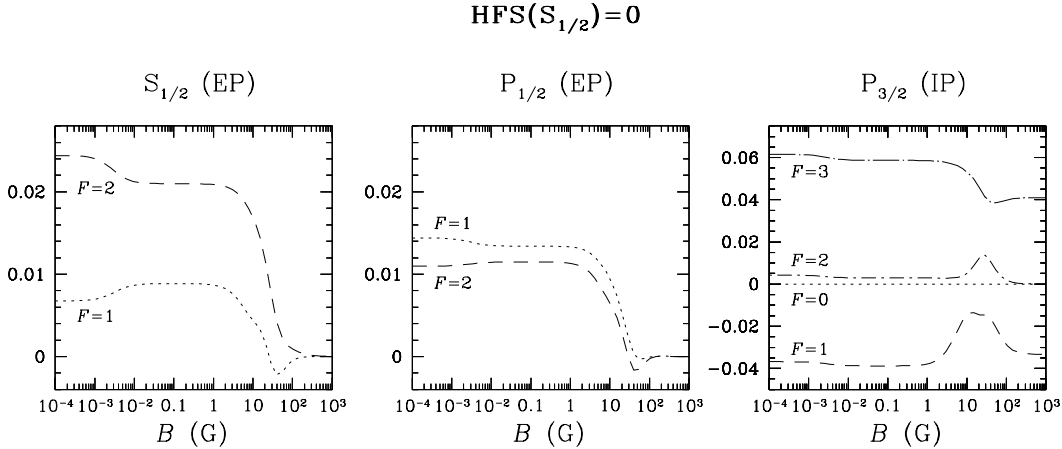}
\figcaption{%
\label{fig:small.HFS}
The same calculations of Figure~\ref{fig:alignment}, but assuming
that no HFS is present in the ground level S$_{1/2}$.
These results show that transfer of atomic alignment from the 
IP level P$_{3/2}$ to the EP level S$_{1/2}$ still occurs,
even if $\vec J$ and $\vec I$ are completely decoupled in the 
EP level, so far that $J$-$I$ coupling is present in the IP level.}
\end{figure*}


\section{Conclusions}
\label{sec:concl}

In this paper we demonstrated analytically that the presence of $J$-$I$
coupling in the IP level P$_{3/2}$ of \ion{Na}{1} is a necessary 
condition for the transfer of atomic alignment 
from that level to the EP ground level S$_{1/2}$. 
We based our demonstration on the quantum 
theory of line formation, as developed by \citet{LA83,LA84,LA85},
and assuming unpolarized incident radiation without spectral 
structure over the frequency intervals 
encompassing the HFS components of the atomic transitions of
interest. Under these conditions, we relied on the argument that 
nuclear polarization cannot be created in an atom 
having only one IP level, if $\vec J$ and $\vec I$ are completely 
decoupled in that level, 
because the assumed incident radiation cannot induce 
directly nuclear transitions in the atom.
It follows, from the results of \S 3, that atomic polarization 
cannot be created in the EP levels, when $\vec J$ 
and $\vec I$ are completely decoupled in the IP level.

We can further strengthen this argument by showing that the 
possibility of nuclear polarization actually resides in the 
presence of $J$-$I$ coupling in the IP level, whereas the 
presence of $J$-$I$ coupling in the EP level is not relevant.
To this purpose, we repeated the calculation of 
Figure~\ref{fig:alignment} after artificially zeroing the HFS 
separation in the level S$_{1/2}$ of \ion{Na}{1}. The results 
of this calculation are shown in Figure~\ref{fig:small.HFS}. Since 
$\vec J$ and $\vec I$ are completely decoupled in the
``modified'' level S$_{1/2}$, the factorization (\ref{eq:rho.fact}) 
always applies to this level. On the other hand, any atomic 
alignment in this modified EP level requires the presence of nuclear
polarization (cf.~eq.~[\ref{eq:rhoKQ.JI}]), since the
electronic angular momentum of the level is $J=1/2$. Such 
nuclear polarization in the EP level S$_{1/2}$ can 
only come from the atomic polarization of the IP level P$_{3/2}$
(which is transferred to the EP level via radiative de-excitation), 
since it is not possible for the prescribed radiation field to 
directly create atomic polarization in the EP level. From the results
of Figure~\ref{fig:small.HFS}, it is evident that the nuclear 
polarization in the modified level S$_{1/2}$ vanishes when the 
regime of complete Paschen-Back effect is reached in the level 
P$_{3/2}$, and eq.~(\ref{eq:rho.fact}) also applies to that 
level. Comparing the results of Figures~\ref{fig:alignment} 
and \ref{fig:small.HFS}, we see that the suppression of 
$J$-$I$ coupling in the level S$_{1/2}$ does not alter 
substantially the SE of the model atom. On the basis of 
these arguments, it seems safe to conclude that, even in
the real case, nuclear polarization cannot be created in the 
atom, when the regime of complete Paschen-Back effect is 
reached in the (only) IP level.

Finally, we must emphasize that the presence of atomic alignment 
in the upper level of the D$_1$ line induces a characteristic
{\em antisymmetric} signature in the core of the Stokes-$Q$ 
profile resulting from the scattering of the anisotropic radiation 
illuminating the atom (see Fig.~2 of Paper I). This applies 
particularly to the optically thin ``prominence case'' 
considered in Paper~I, where the scattering polarization is 
solely due to the emission events following atomic excitation 
by the anisotropic radiation. Currently we are investigating
to what extent such antisymmetric signature can be modified 
through dichroism and radiative transfer effects,
because of the presence of atomic alignment in the ground level
of \ion{Na}{1} (see, e.g., \citealt{TB97}; for the observable 
effects of dichroism and ground-level polarization on the 
\ion{He}{1} 10830 \AA\ multiplet, see \citealt{NATURE}).

In this respect, 
it is interesting to note that spectropolarimetric observations
of the \ion{Na}{1} D-lines obtained with 
TH\'EMIS\footnote{TH\'EMIS is a polarization-free
solar telescope operated by CNRS-CNR in the Spanish Observatorio del Teide
of the Instituto de Astrof\'\i sica de Canarias.} 
in quiet regions close to the solar limb show an antisymmetric
signature in the fractional linear polarization $Q/I$ of the D$_1$ line
(see Fig.~1 of \citealt{TB01b}, which was
adapted from \citealt{TB01}; see also \citealt{BOM02}).
There seems to be an indication of a similar antisymmetric signature
in the $Q/I$ atlas of \citet{GANDO}, which was obtained
with the polarimeter ZIMPOL-II attached to the Gregory Coud\'e
Telescope (GCT) of IRSOL at Locarno (Italy).
On the contrary, analogous observations that
\citet{STEN00} had obtained previously with the polarimeter ZIMPOL-I
attached to the McMath-Pierce facility of the National Solar Observatory 
show almost symmetric profiles
with a central positive peak (see their Fig.~3).\footnote{These 
$Q/I$ observations of quiet solar regions were obtained 
in March 1998 (i.e., two years earlier than the above mentioned 
TH\'EMIS observations), when the Sun had not
yet reached the maximum of its magnetic activity cycle.}

As shown in Paper I, for single-scattering events, one
should expect a {\em symmetric} shape of the Stokes-$Q$ 
signature in the core of the D$_1$ line for magnetic fields 
$B\ga 50\,\rm G$  
(see Fig.~2 of Paper I; note that such symmetric signature 
would change its sign if we considered, say, a horizontal 
canopy-like field instead of the vertical field assumed
for the calculation of that figure). Nevertheless, we think 
that the above mentioned linear-polarization observations of 
the D$_1$ line in very quiet regions of the solar disk,
with TH\'EMIS and ZIMPOL, both have the same physical origin,
i.e., atomic alignment in the levels of the \ion{Na}{1} D$_1$ line.
Now that we understand
how the ground level of \ion{Na}{1} becomes polarized, and how 
its polarization is modified by the presence of weak magnetic 
fields, it will be worthwhile to investigate the sodium 
polarization problem by means of full radiative transfer simulations, 
taking also into account the quantum interferences among the two upper levels 
of the ``enigmatic'' \ion{Na}{1} D-lines.

\acknowledgments

The authors are grateful to Philip Judge and Arturo L\'opez
Ariste (both of HAO), and to Rafael Manso Sainz (Universit\`a di
Firenze, Italy), for reading the manuscript, and for helpful 
comments and suggestions. They also thank Maurizio Landi Degl'Innocenti
(Italian Council for National Research) for helpful discussions about 
the principle of spectroscopic stability during the early stages of 
this work. Thanks are also due to Jan Stenflo and co-workers
for some useful 
discussions and clarifications concerning their spectropolarimetric 
observations. One of the authors (J.T.B.) acknowledges
the support of the Spanish Ministerio de Ciencia y Tecnolog{\'\i}a 
through project AYA2001-1649.


\begin{thebibliography}{}

\bibitem[\protect\citeauthoryear{Bommier \& Molodij}{2002}]{BOM02}
Bommier, V., \& Molodij, G. 2002, \aap, 381, 241.

\bibitem[\protect\citeauthoryear{Kemp, Macek \& Nehring}{1984}]{KE84}
Kemp, J.~C., Macek, J.~H., \& Nehring, F.W.~1984, \apj, 278, 863

\bibitem[\protect\citeauthoryear{Gandorfer}{2000}]{GANDO}
Gandorfer, J.~2000, {\it The Second Solar Spectrum: a high spectral resolution
polarimetric survey of scattering polarization at the solar limb in graphical
representation}, vdf Hochschulverlag AG an der ETH Z\"urich

\bibitem[\protect\citeauthoryear{Landi Degl'Innocenti}{1983}]{LA83}
Landi Degl'Innocenti, E.~1983, \solphys, 85, 3

\bibitem[\protect\citeauthoryear{Landi Degl'Innocenti}{1984}]{LA84}
Landi Degl'Innocenti, E.~1984, \solphys, 91, 1

\bibitem[\protect\citeauthoryear{Landi Degl'Innocenti}{1985}]{LA85}
Landi Degl'Innocenti, E.~1985, \solphys, 102, 1

\bibitem[\protect\citeauthoryear{Landi Degl'Innocenti}{1998}]{LA98}
Landi Degl'Innocenti, E.~1998, \nat, 392, 256

\bibitem[\protect\citeauthoryear{Landolfi \& 
Landi Degl'Innocenti}{1985}]{LL85}
Landolfi M., \& Landi Degl'Innocenti, E.~1985, \solphys, 98, 53

\bibitem[\protect\citeauthoryear{Lehmann}{1969}]{LE69}
Lehmann, J.~C.~1969, Phys.~Rev., 178, 153

\bibitem[\protect\citeauthoryear{Stenflo}{2001}]{STEN01}
Stenflo, J.~O.~2001, in ASP Conf.~Ser.~236, Advanced Solar 
Polarimetry: 
Theory, Observation and Instrumentation, ed.~M.~Sigwarth (San
Francisco: ASP), 97

\bibitem[\protect\citeauthoryear{Stenflo, Gandorfer \& Keller}{2000}]{STEN00}
Stenflo, J.~O., Gandorfer, A. \& Keller, C.~U.~2000, \aap, 355, 781 

\bibitem[\protect\citeauthoryear{Trujillo Bueno}{2001}]{TB01a}
Trujillo Bueno, J.~2001, in ASP Conf.~Ser.~236, Advanced Solar 
Polarimetry: 
Theory, Observation and Instrumentation, ed.~M.~Sigwarth (San
Francisco: ASP), 161

\bibitem[\protect\citeauthoryear{Trujillo Bueno \& Landi
Degl'Innocenti}{1997}]{TB97}
Trujillo Bueno, J., \& 
Landi Degl'Innocenti, E.~1997, \apjl, 482, L183

\bibitem[\protect\citeauthoryear{Trujillo Bueno et al.}{2002a}]{NATURE}
Trujillo Bueno, J., Landi Degl'Innocenti, E.,
Collados, M., Merenda, L. \& Manso Sainz, R.~2002a, \nat, 415, 403

\bibitem[\protect\citeauthoryear{Trujillo Bueno et al.}{2002b}]{TB02}
Trujillo Bueno, J., Casini, R., Landolfi, M., \& 
Landi Degl'Innocenti, E.~2002b, \apjl, 566, L53 (Paper~I)

\bibitem[\protect\citeauthoryear{Trujillo Bueno et al.}{2001}]{TB01}
Trujillo Bueno, J., Collados, M., Paletou, F., \& Molodij, G.~2001, 
in ASP Conf.~Ser.~236, Advanced Solar Polarimetry:
Theory, Observation and Instrumentation, ed.~M.~Sigwarth (San
Francisco: ASP), 141

\bibitem[\protect\citeauthoryear{Trujillo Bueno \& Manso
Sainz}{2001}]{TB01b}
Trujillo Bueno, J., \& Manso Sainz, R.~2001, 
in ASP Conf.~Ser.~248, Magnetic Fields across the Hertzsprung-Russell
diagram, eds.~G.~Mathys, S.K.~Solanki, \& D.T.~Wickramasinghe (San
Francisco: ASP), 83

\bibitem[\protect\citeauthoryear{Trujillo Bueno \& Manso
Sainz}{2002}]{TBMA02}
Trujillo Bueno, J., \& Manso Sainz, R.~2002, Il Nuovo Cimento, in press 

\bibitem[\protect\citeauthoryear{Varshalovich, Moskalev \&
Khersonskii}{1988}]{VA88}
Varshalovich, D.~A., Moskalev, A.~N., \& Khersonskii, V.~K.~1988,
The Quantum Theory of Angular Momentum (Singapore: World Scientific)

\end{thebibliography}
\end{document}